\documentstyle[twocolumn,aps]{revtex}

\begin{document}
\draft
\preprint{HEP/123-qed}

\title{New high-efficiency source of photon pairs for engineering quantum entanglement\\}
\author{Kaoru Sanaka, Karin Kawahara, and Takahiro Kuga}
\address{Institute of Physics, University of Tokyo at Komaba, 3-8-1 Komaba, Meguro-ku, Tokyo 153-8902, Japan\\ and Core Research for Evolutional Science and Technology (CREST), JST, Japan
}

\date{\today}
\maketitle
\begin{abstract}
We have constructed an efficient source of photon pairs using a waveguide-type nonlinear device and performed a two-photon interference experiment with an unbalanced Michelson interferometer. Parametric down-converted photons from the nonlinear device are detected by two detectors located at the output ports of the interferometer. Because the interferometer is constructed with two optical paths of different length, photons from the shorter path arrive at the detector earlier than those from the longer path. We find that the difference of arrival time and the time window of the coincidence counter are important parameters which determine the boundary between the classical and quantum regime. When the time window of the coincidence counter is smaller than the arrival time difference, fringes of high visibility (80$\pm$ 10\%) were observed. This result is only explained by quantum theory and is clear evidence for quantum entanglement of the interferometer's optical paths.
\end{abstract}


\narrowtext

Two-photon entanglement has attracted considerable interest for studying the nonlocal correlations of quantum theory\cite{Einstein,Bell,Clauser,Franson}, and many experiments have been performed\cite{Ou,Shih1,Kwiat,Shih2,Shih3}. The contradiction of local realism can be realized more clearly with multi-photon entanglement systems\cite{GHZ}, which have been demonstrated experimentally in recent years\cite{Pan}. We can expect these systems to be used for novel applications such as quantum cryptography\cite{Ekert}, and quantum teleportation \cite{Bennett}. 

Multi-photon entanglement systems can be generated by parametric down-conversion. Since the probability of generating multi-photon entangled systems decreases exponentially with the number of entangled photons, it becomes more difficult to conduct experiments with a large number of entangled photons\cite{Sackett}. One of the candidates for solving this difficulty is to make the ultrabright source of polarization-entangled photons proposed by Kwiat {\em et al.} \cite{Kwiat2}. The source is superior to other sources because nearly every pair of photons is polarization entangled. Since the total number of generated photon pairs is limited by the nonlinear susceptibility and phase matching condition of a nonlinear crystal, a remarkable increase in the number of photon pairs can not be expected if one uses bulk crystals. This is to be compared to the drastic improvement of the efficiency to generate photon pairs we present. Our method uses a waveguide type nonlinear device originally developed for type-I quasi-phase-matching frequency doubling. Using the newly developed source of photon pairs, we then perform a two-photon interference experiment and show that photon pairs are in the entangled state for interferometer's optical paths. Parametric down-converted photons from the nonlinear device are detected by two detectors located at the output ports of the interferometer. Because this interferometer is constructed with two optical paths of different length, photons from the shorter path arrive at the detector earlier than those from the longer path. When the time window of the coincidence counter is larger than the arrival time difference, in other words when one can not distinguish the optical paths, the wave function that causes the coincidence is the superposition of all possible state. We define the states to be; $\psi (S, S)$, $\psi (L, L)$, $\psi (S, L)$, and $\psi (L, S)$, where $\psi (S, S)$ ($\psi (L, L)$) corresponds to the photons both passing along the longer (shorter) path and $\psi (S, L)$ and $\psi (L, S)$ corresponds to one photon passing along the longer path and another passing along the shorter path of the interferometer. On the other hand, when the time window of coincidence counter is smaller than the arrival time difference, the wave function is the superposition of $\psi (S, S)$ and $\psi (L, L)$, i.e. the entanglement of optical paths. Quantum theory predicts that the state of entangled optical paths should have high visibility two-photon interference fringes with contrasts over 50\%. We have observed a maximum visibility of 80 $\pm$ 10\% in the experiment, a clear evidence for entanglement.

The schematic of our experimental setup is shown in Fig.\ \ref{illustration}. Photon pairs generated by parametric down-conversion are injected collinearly into one input port of an unbalanced Michelson interferometer which is constructed with optical paths $S$ and $L$. The coincidence measurement between the two outputs of the interferometer shows two-photon interference.
The two-photon state in the process of spontaneous parametric down-conversion is described as
\begin{equation}
\mid\psi_i \rangle=\int d k_1 \int d  k_2 \  \delta (k_p-k_1-k_2) \ \Phi (k_1)  \mid k_1 \rangle \otimes  \mid k_2 \rangle,
\label{2photonstate-i}
\end{equation}
where $k_1$, $k_2$, and $k_p$ are signal, idler, and pump wave number, respectively. The $\delta$ function comes from the perfect phase-matching condition of the parametric down-conversion, $\Phi (k)$ is the wave-packet distribution function, and its width $\Delta k$ determines the coherence length of the down-conversion field as $\ell_{coh}=1/\Delta k$. Here the optical-path difference $\Delta L=L-S$ satisfies the condition
\begin{equation}
\Delta L>>\ell_{coh} \ .
\label{coherent}
\end{equation}
So that single-photon interference effects are negligible. After passing through the interferometer, the two-photon state becomes
\begin{equation}
\mid\psi_f \rangle=\int d k_1 \int d  k_2 \  \delta (k_p-k_1-k_2) \ \Phi (k_1)  \mid k_A \rangle \otimes  \mid k_B \rangle,
\label{2photonstate-f}
\end{equation}
where $\mid k_A \rangle$ and $\mid k_B \rangle $ represent photon states at detectors $D_A$ and  $D_B$. These states are described by
{\setcounter{enumi}{\value{equation}}
\addtocounter{enumi}{1}
\setcounter{equation}{0}
\renewcommand{\theequation}{\theenumi\alph{equation}}
\begin{eqnarray}
\mid k_A \rangle &=& \frac{1}{2} \  [ \mid k_{1 S} \rangle+  \mid k_{1 L} \rangle+  \mid k_{2 S} \rangle+  \mid k_{2 L} \rangle  ], \label{superposition1}\\
\mid k_B \rangle &=& \frac{1}{2} \ [ \mid k_{1 S} \rangle-  \mid k_{1 L} \rangle+ \mid k_{2 S} \rangle-  \mid k_{2 L} \rangle ],\label{superposition2}
\end{eqnarray}
\setcounter{equation}{\value{enumi}}
}\\
\\
where $\mid k_{n l} \rangle = \mid k_n \rangle e^{i k_n l}$. Substituting (\ref{superposition1}) and (\ref{superposition2}) into (\ref{2photonstate-f}), we obtain 
\begin{eqnarray}
& & \mid\psi_f \rangle = \frac{1}{4}@\int d k_1 \int d  k_2 \ \delta (k_p-k_1-k_2) \ \Phi (k_1)  \times \nonumber \\ 
& &[\mid k_{1 S} \rangle \mid k_{2 S} \rangle + \mid k_{2 S} \rangle \mid k_{1 S} \rangle - \mid k_{1 L} \rangle \mid k_{2 L} \rangle - \mid k_{2 L} \rangle \mid k_{1 L} \rangle \nonumber \\
&-&  \mid k_{1 S} \rangle \mid k_{2 L} \rangle - \mid k_{2 S} \rangle \mid k_{1 L} \rangle + \mid k_{1 L} \rangle \mid k_{2 S} \rangle + \mid k_{2 L} \rangle \mid k_{1 S} \rangle ]. \nonumber \\
& &  \nonumber \\
& &
\label{general}
\end{eqnarray}
\\
For example, $\mid k_{1 S} \rangle \mid k_{2 S} \rangle$ and $\mid k_{2 S} \rangle \mid k_{1 S} \rangle$ in (\ref{general}) correspond to the photons which have followed the $(S, S)$ path in the interferometers. The coincidence rate can be estimated to be $R_c=R_{c0} \langle \psi_f \mid \psi_f \rangle$,
\begin{eqnarray}
&R_c& = \frac{R_{c0}}{2} \int d k_1  \ \mid \Phi (k_1) \mid ^{2} \nonumber \\ 
& \times & [1- \frac{1}{2} \cos k_p \Delta L- \frac{1}{2} \cos (k_p - 2 k_1) \Delta L ], \nonumber \\ 
& &
\label{coincidence}
\end{eqnarray}
when
\begin{equation}
\Delta T >\Delta L / c ,
\label{window1}
\end{equation}
where $\Delta T$ is the time window of coincidence counter. Because $\Delta L$ is greater than the first-order coherence length of the wave packets, the last term in (\ref{coincidence}) will vanish and we have,
\begin{eqnarray}
&R_c& \simeq \frac{R_{c0}}{2} \int d k_1  \ \mid \Phi (k_1) \mid ^{2} \  [1- \frac{1}{2} \cos k_p \Delta L ]. \nonumber \\
& &
\label{coincidence2}
\end{eqnarray}
A similar result can be derived from a classical model. The wave number $k_1$ and $k_2$ are classical random variables which are subject to the constraint that $k_p=k_1+k_2$, where $k_p$ is a nonrandom variable.
\begin{eqnarray}
R_1 &\propto& \langle 1 + \cos k_1 \Delta L \rangle \simeq 1, \nonumber \\ 
R_2 &\propto& \langle 1 - \cos k_2 \Delta L \rangle \simeq 1, \nonumber \\
R_c &\propto& \langle (1 + \cos k_1 \Delta L)(1 - \cos k_2 \Delta L) \rangle \nonumber \\
&\simeq& 1 - \frac{1}{2} \cos k_p \Delta L. \nonumber \\
& &
\label{classical}
\end{eqnarray}
Both quantum (\ref{coincidence2}) and classical (\ref{classical}) models predict 50\% visibility.

On the other hand, when the time window of coincidence counter $\Delta T$ is small enough to distinguish each photon of the pair coming from different paths
\begin{equation}
\Delta T <\Delta L / c ,
\label{window2}
\end{equation}
the last four terms of (\ref{general}) are not be registered by the coincidence counter. In this case, the wave function that causes the coincidence becomes
\begin{eqnarray}
& & \mid\psi_f \rangle = \frac{1}{4}@\int d k_1 \int d  k_2 \ \delta (k_p-k_1-k_2) \ \Phi (k_1)  \times \nonumber \\ 
& &[\mid k_{1 S} \rangle \mid k_{2 S} \rangle + \mid k_{2 S} \rangle \mid k_{1 S} \rangle - \mid k_{1 L} \rangle \mid k_{2 L} \rangle - \mid k_{2 L} \rangle \mid k_{1 L} \rangle ]. \nonumber \\
& &  \nonumber \\
& &
\label{nallow}
\end{eqnarray}
\\

Here the quantum theoretical calculation predicts the coincidence rate to be
\begin{eqnarray}
&R_c& = \frac{R_{c0}}{4} \int d k_1  \ \mid \Phi (k_1) \mid ^{2} \  [1- \cos k_p \Delta L ], \nonumber \\
& &
\label{coincidence:nallow}
\end{eqnarray}
and we expect fringes with 100\% visibility. Two-photon interference fringes with over 50\% visibility can never be explained with classical models\cite{Shih2,Shih3}.

Under a quasi-monochromatic wave model $ k \simeq k_1 \simeq k_2$, (\ref{nallow}) becomes
\begin{eqnarray}
\mid\psi_{entangle} \rangle &=& \frac{1}{2} \int d k  \ \delta (k_p- 2 k) \ \Phi (k)\nonumber \\ 
& \times &[\mid k_S \rangle \mid k_S \rangle - \mid k_L \rangle \mid k_L \rangle ]. 
\label{entangle}
\end{eqnarray}
\\
This means that two-photon interference with over 50\% visibility reflects the  two-photon entangled state of the interferometer's optical paths. 

In the experimental arrangement, we utilize two waveguide type nonlinear devices fabricated on a 1-cm-long LiNbO$_3$ substrate, one for frequency doubling and another for generating photon pairs. A CW laser beam (854 nm, 10 mW) from a distributed Bragg reflector (DBR) laser diode (SDL-5702-H1) is converted to violet light (427 nm) by type-I quasi-phase-matching second-harmonic-generation (SHG). Due to the high conversion efficiency over 1 \%, we can obtain more than 0.1 mW of violet light\cite{Matsushita}. After passing through a pellin broca prism and a blue filter (BF), the violet light is sent to the second device and conjugated photon pairs around 854 nm wavelength are generated in the process of spontaneous parametric down-conversion (PDC). We estimate that $10^5$- s$^{-1}$ photon pairs are generated with this weak violet light when we take into account the detection efficiency. A low pass filter (LPF) and dichroic mirrors are used to separate out the violet beam.

The collinear signal and idler photon pairs are injected into an input port of Michelson interferometer composed of a 50\%-50\% non-polarized beam splitter (BS) and retroreflectors. The optical-path difference of the interferometer is arranged about 55-cm to satisfy (\ref{coherent}), and can be moved by piezoelectric ceramic actuator (PZT). Two beams from the output ports of the interferometer are fed into single photon detectors $D_A$ and $D_B ($EG\&G SPCM-AQR14). One signal is used for the start signal of a time to amplitude converter (TAC) and the other is used for the stop signal after it passes through an electrical delay line. We record the pulse height distribution with a multi-channel analyzer (MCA) for 10 s, under computer control (PC).

An example of a pulse height distribution obtained with this interferometer is shown in Fig.\ \ref{distribution}. There are three distinct peaks in the figure and they correspond to photon pairs coming through $(L, S)$, $(S, S)$ or $(L, L)$, and $(S, L)$ optical paths from left to right, respectively. Because we record the time interval distribution of coincidence counts, we can analyze two-photon interference visibility with delayed choice of the coincidence time window. In the case where we are summing up all of three peaks, i.e., $\Delta T$ = 5 ns and the experimental condition satisfies (\ref{window1}), the visibility of observed fringes should be less than 50\% (classical regime). Fig.\ \ref{int5ns} shows coincidence count rates as a function of the optical-path difference $\Delta L$, while the single count rate is almost constant: 10000 $\pm$ 500 s$^{-1}$. The observed interference visibility of 41 $\pm$ 10 \% is explained by the classical theory. On the other hand, when we sum up only central peak from the same records, i.e., $\Delta T$ = 1 ns and this condition satisfies (\ref{window2}), the visibility of observed fringes could be more than 50\% (quantum regime). The results are shown in Fig.\ \ref{int1ns}. The observed visibility is quite high, 80 $\pm$ 10 \%. Because the visibility is over 50 \%, these results can only be explained by quantum theory, thus proving that the two-photon optical-path entangled state was created. 

We have constructed an efficient source of photon pairs using a waveguide-type nonlinear device. The efficiency of parametric down-conversion is at the same level of that is obtained by bulk nonlinear crystals with about a thousand time greater pump beam \cite{Shih1}. We performed a two-photon interference experiment with the source of photon pairs and an unbalanced Michelson interferometer. When we sum up the region where $ \Delta T > \Delta L / c $ of the record measured by a TAC, the fringe visibility is smaller than 50\%, which can be explained by a classical model. On the other hand, when we sum up only the region of $ \Delta T <\Delta L / c $ from the same records, the observed visibility is 80 $\pm$ 10 \% and clearly exceeds the classical prediction (50\%). These results can only be explained by quantum theory, and clear evidence for quantum entanglement of the interferometer's optical paths.

We can also construct an efficient source of polarization entangled photon pairs using two waveguide type nonlinear devices\cite{Kwiat2}. The high efficiency of this source can be used for experiments that require a lot of photon pairs (quantum cryptography and quantum teleportation), and makes it possible to more efficiently generate multi-photon entanglement which should lead to progress in quantum information technology. 

We are grateful to the members of the Optical Disk Systems Development Center at Matsushita Electric Industrial Co. for their experimental corporation. This work was supported by Matsuo Foundation and Research Foundation for Opto-Science and Technology.

\newpage

\begin{figure}
\caption{Schematic of the two-photon interference  experiment in an unbalanced Michelson interferometer. The path difference is set large enough such that the single-photon interference effect is nearly equal to zero. }
\label{illustration}
\end{figure}

\begin{figure}
\caption{Measured time difference distributions for optical-path difference of the interferometer.}
\label{distribution}
\end{figure}

\begin{figure}
\caption{Second-order interference fringes with 5-ns coincidence time window. (1V = 46$\pm$ 8 \ nm)}
\label{int5ns}
\end{figure}

\begin{figure}
\caption{Second-order interference fringes with 1-ns coincidence time window. (1V = 46$\pm$ 8 \ nm)}
\label{int1ns}
\end{figure}

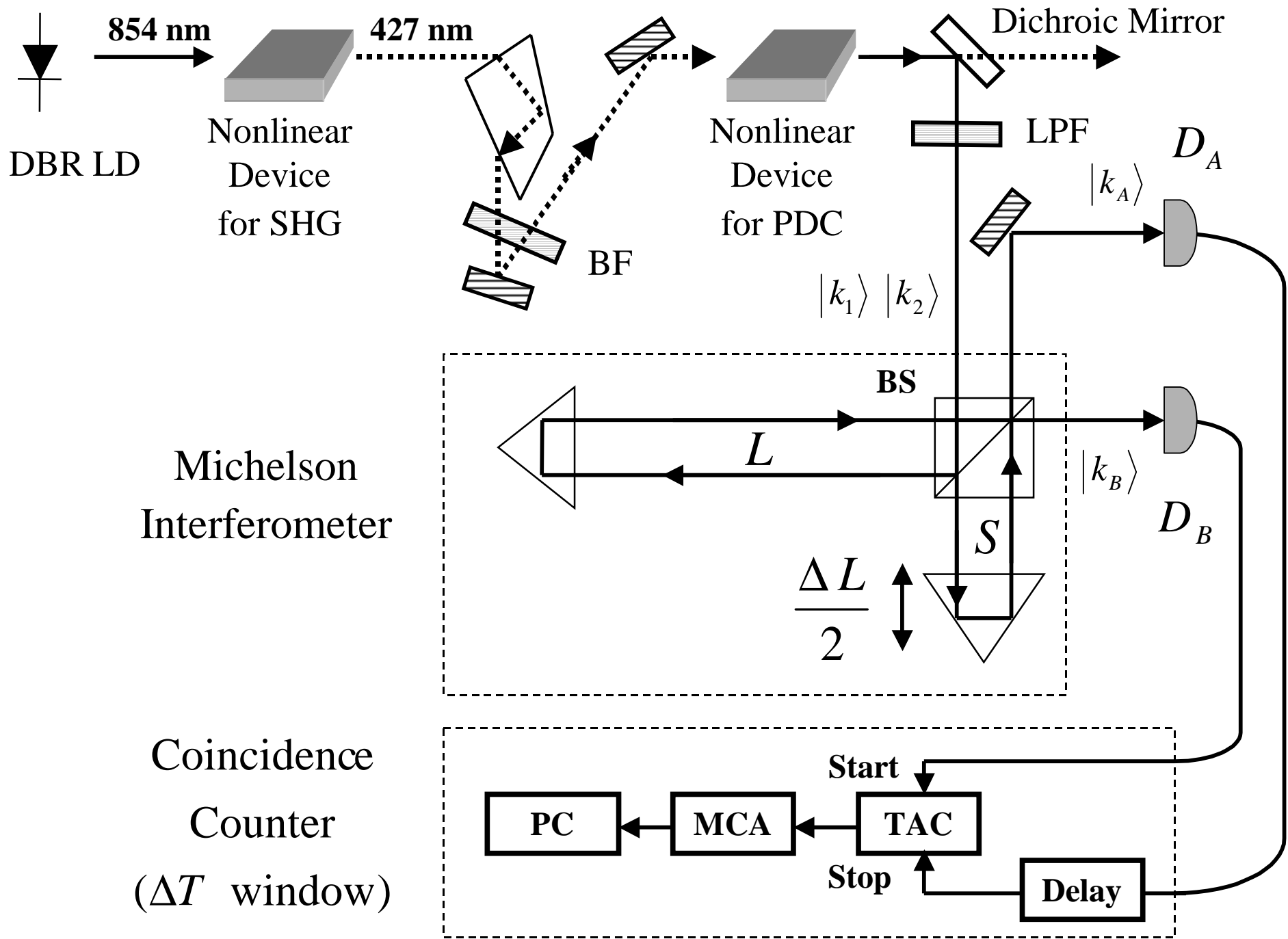
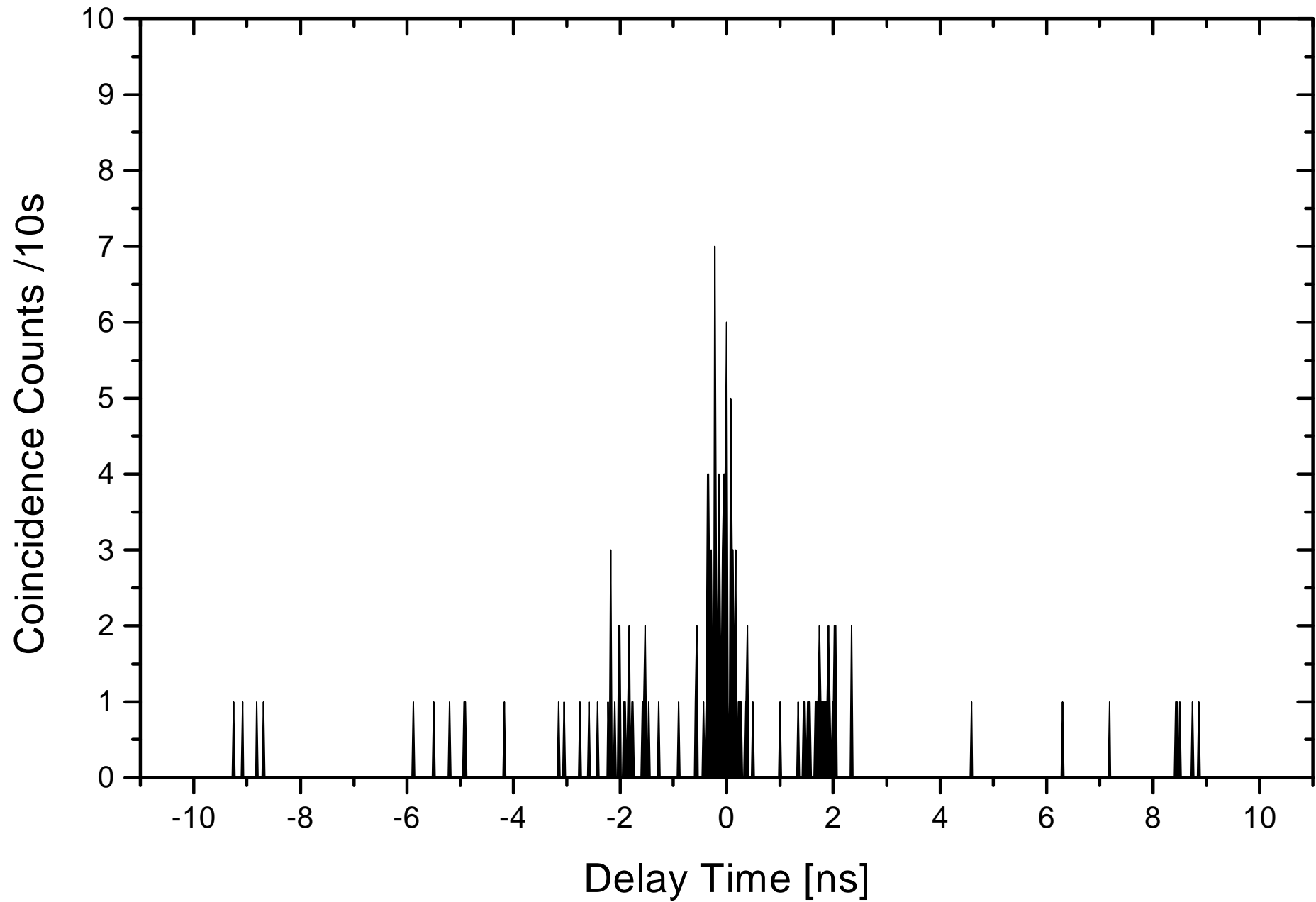
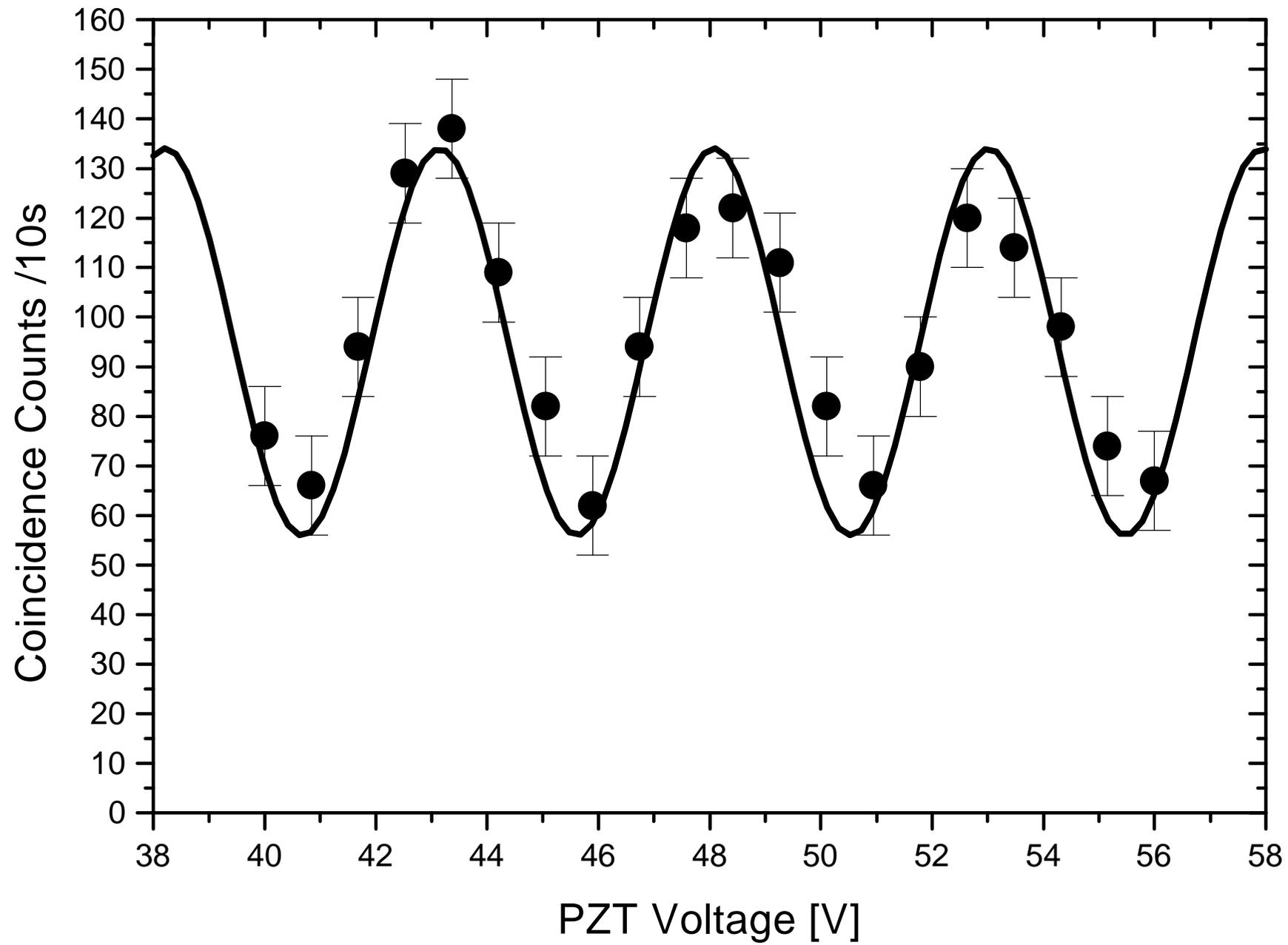
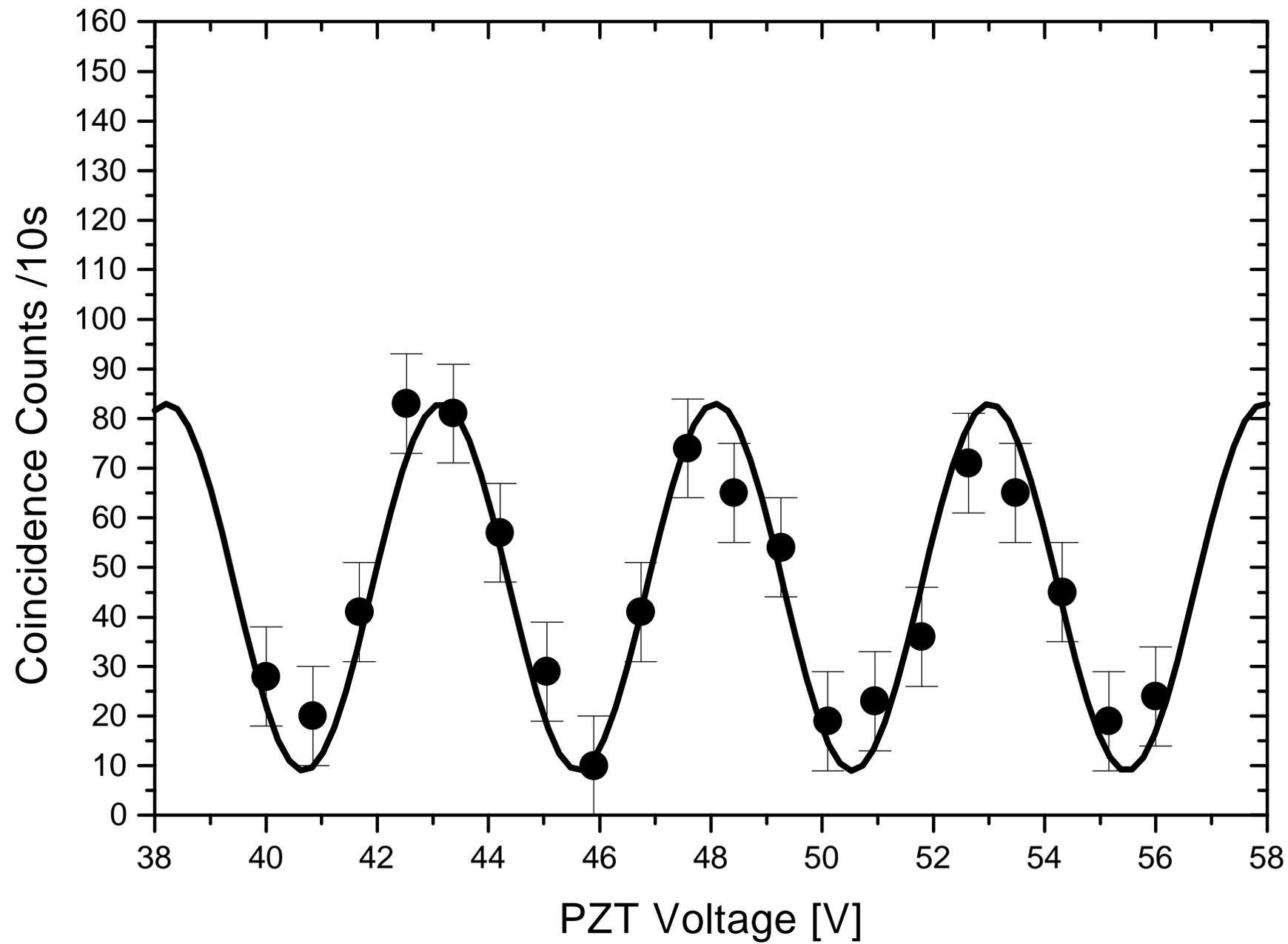
\end{document}